\documentclass[11pt,a4paper,twocolumn]{revtex4-2}
\selectfont
\usepackage[
  top=20mm,
  bottom=20mm,
  left=14mm,
  right=14mm,
]{geometry}
\usepackage{amsmath,amssymb,amsthm,bm,mathrsfs,mathtools}
\usepackage[utf8]{inputenc}
\usepackage[T1]{fontenc}
\usepackage{graphicx}
\usepackage[dvipsnames]{xcolor}
\usepackage{hyperref}
\usepackage{siunitx}
\usepackage{fmtcount}
\usepackage{textcomp}
\usepackage{flafter}
\usepackage{caption}      
\usepackage{fancybox}    
\usepackage{mdframed}   
\usepackage{tcolorbox}

\date{\today}

\begin{document}
\title{Secret Entanglement, Public Geometry.\\
Quantum Cryptography from a Geometric Perspective}
\author{Loris Di Cairano}
\email{l.di.cairano.92@gmail.com, loris.dicairano@uni.lu}

\affiliation{Department of Physics and Materials Science, University of Luxembourg, L-1511 Luxembourg City, Luxembourg}

% We explore a simple idea: use the geometry of projective Hilbert space and its foliation by entanglement to define elementary cryptographic schemes.
% The basic ingredients are \textit{(i)} the Fubini-Study metric on the space of pure states, \textit{(ii)} a choice of an entanglement measure viewed as a function on this manifold, and \textit{(iii)} controlled trajectories generated by unitary operations.
% The protocol is public at the level of the geometric framework and the encoding map, but it depends on a secret choice of entanglement measure (or, more precisely, on hidden parameters that define a foliation into constant-entanglement hypersurfaces).
% We formalize this idea in terms of \textit{geometric entanglement codes} and illustrate it with a toy example of two qubits, where the key is a local unitary that twists the foliation.
% Our goal is to propose a concrete geometric framework that can serve as a playground for more systematic developments.
\begin{abstract}
Can a secret be hidden not in which quantum state is prepared, but in the way that state \emph{moves} through its space of possibilities? Motivated by this question, we propose an essential geometric perspective on quantum cryptography in which projective Hilbert space and its entanglement foliations play a central role. The basic ingredients are: (a) the Fubini-Study metric on the manifold of pure states, (b) a family of entanglement measures viewed as scalar functions on this manifold, and (c) controlled trajectories generated by unitary operations.

The geometric structure—state manifold, metric, and allowed moves—is fully public, as is the functional \emph{form} of the entanglement family. What remains secret is the choice of parameter $\theta$ that selects a specific entanglement functional $E_\theta$ and the corresponding foliation into constant-entanglement hypersurfaces. In this setting, classical messages are encoded not only in the sequence of states but also in the pattern of upward, downward, or tangential steps with respect to the hidden foliation. We formalize this idea in terms of \emph{geometric entanglement codes} and illustrate it with two toy constructions in which incompatible foliations play the role of mutually unbiased bases. 
\end{abstract}

\maketitle

\section{State of the art}

Quantum cryptography is one of the most mature areas of quantum information science. 
Since the original BB84 proposal by Bennett and Brassard~\cite{BennettBrassard84,BennettBrassard84TCS}, quantum key distribution (QKD) has evolved from a conceptual curiosity into a full-fledged research field with commercial prototypes and long-distance field trials. 
In its simplest prepare–and–measure form, QKD exploits the fact that non-orthogonal quantum states cannot be measured without disturbance, allowing two parties to establish a secret key with information-theoretic security, independent of any computational assumption~\cite{GisinRMP02,LoCurtyTamaki14,XuMaZhangLoPan20,PirandolaAOP20}.
The basic ideas of BB84 have been generalized in many directions, e.g., the B92 and six-state protocols~\cite{Bennett92B92,Bruss98SixState}, 
and underlie most modern discrete-variable QKD systems.

Shortly after BB84, Ekert showed that entanglement and Bell inequalities provide an alternative, conceptually appealing route to QKD~\cite{Ekert91}.
In his protocol (E91), security is certified by the violation of a Bell inequality rather than by the disturbance of non-orthogonal states, leading to entanglement-based schemes such as BBM92 and to early ideas connecting QKD and entanglement distillation~\cite{BBM92,LoChau99,ShorPreskill00}.
These developments laid the foundations for device-independent QKD, where security is derived almost solely from the observed statistics and the validity of quantum theory, with minimal assumptions on the internal workings of the devices~\cite{Acin07DIQKD,Pironio09DIQKD,VaziraniVidick14,Zapatero23DIReview}.
A large body of work has since developed around entanglement-based and device-independent protocols, including composable security proofs and increasingly sophisticated experiments bridging theory and realistic implementations~\cite{GisinRMP02,ScaraniRMP09,LoCurtyTamaki14,XuMaZhangLoPan20,PirandolaAOP20}.

Over the last two decades, the QKD landscape has been thoroughly mapped in several influential reviews. 
Gisin and co-workers provided an early comprehensive overview that emphasized both theoretical principles and experimental progress~\cite{GisinRMP02}, 
while Scarani \emph{et al.}\ focused on the security of practical implementations, including discrete-variable, continuous-variable, and distributed-phase-reference protocols~\cite{ScaraniRMP09}.
Then the state of secure QKD has been reviewed with an emphasis on security models and assumptions~\cite{LoCurtyTamaki14},
and then QKD has been analyzed with realistic devices, side channels, and countermeasures~\cite{XuMaZhangLoPan20}.
For a reader interested in an extensive state-of-the-art survey that covers discrete- and continuous-variable QKD, device-independent schemes, satellite and fiber implementations, and quantum networks, we refer to Ref.~\cite{PirandolaAOP20}.
These works set the standard picture of quantum cryptography: protocols are specified in terms of signal states, measurements and classical post-processing, and their security is analyzed within information-theoretic and composable frameworks.

Within this mainstream development, there are strands that make the geometry of state space more explicit.
One is the use of \emph{spherical codes} for qubits: Renes showed that equiangular spherical codes on the Bloch sphere (such as trine and tetrahedral configurations) can be used to define QKD protocols with advantageous noise tolerance and flexible operating points~\cite{Renes04SphericalCodes,Renes05Equiangular}.
Here, the signal states are literally points on a sphere, and the geometry of their arrangement controls both the information rate and the security properties.
Another strand is high-dimensional QKD, where information is encoded in larger Hilbert spaces, for instance, using mutually unbiased bases in dimension $d>2$ or orbital-angular-momentum (OAM) modes~\cite{BechmannPasquinucciPeres00,CerfBourennaneKarlssonGisin02,Durt10MUBReview,Mafu13OAMQKD}.
In such schemes, high-dimensional constellations increase key rates and robustness, but they also bring to the forefront the structure of the underlying projective Hilbert space, the geometry of state discrimination, and the behavior of error patterns in $d$-level systems.
These ideas are now explored in a wide range of experimental platforms, from OAM-based photonic systems to time-phase encoding and integrated-optics implementations~\cite{Mafu13OAMQKD,Erhard18ReviewHighDim,Bouchard18NPJQI,Paesani19NatPhys}.

Parallel to the cryptographic literature, there is a well-developed body of work that studies the geometry of quantum states for its own sake, with entanglement emerging as a geometric feature of projective Hilbert space.
The monograph by Bengtsson and Życzkowski and related articles develops a detailed picture of the convex and differential geometry of quantum state space, including complex projective spaces, distinguishability metrics, and the stratification of state space into subsets of separable and entangled states~\cite{BengtssonZyczkowskiBook,BengtssonZyczkowskiIntro}. In parallel, the review by Horodecki \emph{et al.} provides a systematic account of entanglement measures, mixed-state entanglement, and its operational role in quantum information~\cite{HorodeckiRMP09}.

Other works draw direct links between the Fubini-Study (FS) metric, local unitary orbits, and entanglement monotones, providing explicit geometric characterizations of how entanglement sits inside $\mathbb{CP}^{d-1}$~\cite{BrodyHughston01,AnandanAharonov90,BraunsteinCaves94,PetzSudar96,Aniello11GeomQM}.
Within this geometric viewpoint, entanglement distances and correlation measures based on the FS metric have been introduced for multipartite and hybrid $M$-qudit systems, mixed multipartite states, and graph-based architectures, offering concrete geometric quantifiers tailored to realistic quantum networks and directed-graph models~\cite{Cocchiarella20EntanglementDistance,Vesperini23SciRep,VesperiniFranzosi24AQT,DeSimoneFranzosi25AQT,DeSimoneFranzosi25JPA}.

In geometric quantum mechanics, projective Hilbert space is treated as a Kähler manifold on which quantum evolutions are curves, observables are functions generating Hamiltonian flows, and entanglement can be interpreted in terms of the geometry of orbits and foliations under local unitary actions~\cite{AshtekarSchilling99,Gibbons92,ChruscinskiJamiolkowskiBook04}.
Recent work has extended this program to define geometric measures of entanglement and entropic functionals directly on projective space, including the geometric entanglement entropy built from the volume of level sets of suitable functionals~\cite{VedralRMP02,OsborneNielsen02,DiCairanoGeoEntEntropy}.

\section{Encoding information in the geometry of quantum states: motivation and idea of our perspective}

Hilbert space is usually treated as an abstract linear space, whose role is to host state vectors and operators.
However, once we mod out global phases, the physically relevant space of pure states is a curved manifold: the projective Hilbert space, equipped with the Fubini-Study (FS) metric.
On this manifold, the geometry encodes nontrivial physical information: distances, volumes, and curvatures can be related to distinguishability, uncertainty, and even entanglement.
In parallel, modern quantum cryptography typically relies on operational primitives such as no-cloning, uncertainty relations, and the impossibility of perfectly distinguishing non-orthogonal states.
The geometric structure of state space is often implicit in these protocols, but it rarely plays a central role in their design.

The present work takes inspiration from both lines: from quantum cryptography, in that we consider schemes where classical information is encoded into quantum states and recovered with the help of a shared secret; and from geometric approaches, in that we treat projective Hilbert space as a Riemannian manifold foliated by constant-entanglement hypersurfaces.
At the kinematic level, the projective space $(\mathcal{M},g_{\text{FS}})$, endowed with the FS metric and its unitary invariance, is fully public and plays the role of a common state manifold, in the spirit of geometric formulations of quantum mechanics and entanglement~\cite{BengtssonZyczkowskiBook,BengtssonZyczkowskiIntro,Aniello11GeomQM}.
In addition, it is publicly known that Alice and Bob probe this manifold through a \emph{family} of entanglement functionals $\{E_\theta\}_{\theta\in\Theta}$, compatible with standard bipartite settings and entanglement-based views of quantum communication and cryptography~\cite{Ekert91,GisinRMP02,ScaraniRMP09,LoCurtyTamaki14,PirandolaAOP20,DiCairanoGeoEntEntropy}.

Concretely, we consider:
\begin{itemize}
  \item the projective space $\mathcal{M} = \mathbb{P}(\mathcal{H})$ of pure states for a finite-dimensional Hilbert space $\mathcal{H}$, endowed with the FS metric $g_{\text{FS}}$;
  \item a family of entanglement measures $E_\theta : \mathcal{M} \to \mathbb{R}$, labelled by parameters $\theta$;
  \item the foliation of $\mathcal{M}$ into constant-entanglement hypersurfaces $\Sigma_e^{(\theta)} = \{[\psi] \in \mathcal{M} : E_\theta([\psi])=e\}$;
  \item trajectories $\gamma(t)$ generated by sequences of unitary operations, which move a state across these leaves.
\end{itemize}

What remains secret is the specific choice of parameter $\theta$ and therefore the concrete entanglement functional $E_\theta$ and its associated foliation $\mathcal{F}_\theta = \{\Sigma_e^{(\theta)}\}$ into constant-entanglement hypersurfaces.
In this sense, the \textit{kinematic geometry} of projective Hilbert space is public, while the \textit{entanglement geometry} relevant for decoding—which leaves matter, how they are ordered in ``height'', and how trajectories are interpreted as going ``up'', ``down'', or tangentially with respect to $E_\theta$—is part of the key.
The key idea is that the \emph{public} part of the protocol fixes the state manifold $(\mathcal{M},g_{\text{FS}})$ and the general form of the family of entanglement functionals $\{E_\theta\}_{\theta\in\Theta}$ that may be used to probe it, whereas the \emph{secret} part specifies which entanglement functional is actually used and how its leaves (level sets) are to be read.

Classical messages are then encoded not only in which state is prepared but also in how a trajectory moves across this hidden foliation: in the pattern of upward, downward, or tangential steps with respect to $E_\theta$ along a sequence of unitary operations.
From the point of view of an observer who does not know $\theta$, the same trajectory may look structureless or even entirely separable, much like in standard QKD, where an eavesdropper who chooses the wrong basis sees only random outcomes~\cite{BennettBrassard84,Bennett92B92,Renes04SphericalCodes,CerfBourennaneKarlssonGisin02}.
In this article, we combine the two viewpoints and ask a simple question:
\begin{quote}
    \emph{Can one design a cryptographic scheme in which the message is encoded not just in ``which state'' is prepared, but also in ``how one moves'' across projective Hilbert space, as seen through a foliation by entanglement?}
\end{quote}
Our aim here is essential: we do not attempt to build a practically secure protocol at this stage but rather to provide a clean geometric construction that can be analyzed and generalized.
We formalize this as a \emph{geometric entanglement code} and then present a minimal two-qubit example, where the key is a local unitary that twists the measure of entanglement.

\section{Geometric preliminaries}

\subsection{Projective Hilbert space and Fubini-Study metric}

Let $\mathcal{H}$ be a finite-dimensional Hilbert space of dimension $d$.
Pure quantum states are rays in $\mathcal{H}$, i.e.\ equivalence classes $[\psi] = \{\lambda|\psi\rangle : \lambda \in \mathbb{C}\setminus\{0\}\}$, with $|\psi\rangle$ normalized.
The set of rays is the complex projective space
\begin{equation}
  \mathcal{M} = \mathbb{P}(\mathcal{H}) \simeq \mathbb{CP}^{d-1}.
\end{equation}
On $\mathcal{M}$, one has a canonical Riemannian metric, the Fubini-Study metric (FS)~\cite{Fubini1904,Study1905,KobayashiNomizuII,BengtssonZyczkowskiBook}.
In a local chart, for normalized representatives $|\psi\rangle$ and tangent vectors $|\delta\psi\rangle$ orthogonal to $|\psi\rangle$, the line element reads
\begin{equation}
  ds^2 = g_{\text{FS}}(\delta\psi,\delta\psi) = 
  \langle \delta\psi|\delta\psi\rangle 
  - |\langle \psi|\delta\psi\rangle|^2.
\end{equation}
The associated distance $d_{\text{FS}}([\psi],[\phi])$ quantifies the distinguishability of pure states; the associated volume form $d\mu_{\text{FS}}$ is unitarily invariant~\cite{Kibble1979,ProvostVallee1980}.

In what follows we will treat $(\mathcal{M},g_{\text{FS}})$ as the ``state manifold'' on which information is encoded geometrically.
Curves $\gamma:[0,1]\to\mathcal{M}$ represent continuous transformations of the state, typically generated by unitary operations.

\subsection{Entanglement as a function on projective space}

Suppose that the Hilbert space factorizes as
\begin{equation}
  \mathcal{H} = \mathcal{H}_A \otimes \mathcal{H}_B,
\end{equation}
with $\dim\mathcal{H}_A = d_A$, $\dim\mathcal{H}_B = d_B$, and $d = d_A d_B$.
For a pure state $|\psi\rangle \in \mathcal{H}$, entanglement between $A$ and $B$ is usually quantified by a scalar function of the reduced density matrix $\rho_A = \mathrm{Tr}_B|\psi\rangle\langle\psi|$.
A standard example is the von Neumann entropy~\cite{vonNeumann1932,NielsenChuangBook}:
\begin{equation}
  E_{\text{vN}}([\psi]) = -\mathrm{Tr}\left(\rho_A \log \rho_A\right),
\end{equation}
which vanishes on product states and is maximal on maximally entangled states.

More generally, we consider a family of entanglement measures
\begin{equation}
  E_\theta : \mathcal{M}\to\mathbb{R}, \qquad \theta\in\Theta,
\end{equation}
where $\Theta$ is some parameter space.
Now, we do not specify any particular family but only require that:
\begin{itemize}
  \item each $E_\theta$ is smooth on $\mathcal{M}$ (except possibly on a set of measure zero);
  \vspace{-0.2cm}
  \item local unitary transformations (on $A$ and $B$ separately) act as isometries of $(\mathcal{M},g_{\text{FS}})$ and, for each fixed $\theta$, preserve the level sets of $E_\theta$ up to reparametrization.
\end{itemize}

\subsection{Geometric construction of the entanglement level sets}\label{ssec:geometric-constr}

Given an entanglement functional $E_\theta:\mathcal{M}\to\mathbb{R}$, we regard it as a height function on the projective manifold $(\mathcal{M},g_{\text{FS}})$.
The differential $dE_\theta$ is a one-form on $\mathcal{M}$, and using the FS metric, we can associate a unique gradient vector field $\nabla E_\theta$ defined by~\cite{LeeSmooth}
\begin{equation}
  g_{\text{FS}}(\nabla E_\theta(x), v)
  \;=\; dE_\theta(x)[v]
  \qquad \text{for all } v\in T_x\mathcal{M}.
  \label{eq:grad-def}
\end{equation}
In local coordinates, this is the usual Riemannian gradient
\begin{equation}
  (\nabla E_\theta)^i
  \;=\; g_{\text{FS}}^{ij}\,\partial_j E_\theta,
\end{equation}
where $g_{\text{FS}}^{ij}$ is the inverse of the FS metric tensor $g^{\text{FS}}_{ij}$. The integral curves of $\nabla E_\theta$ are those that increase $E_\theta$ most rapidly with respect to the FS distance. The leaf or level set
\begin{equation}
  \Sigma_e^{(\theta)} := \{\, x\in\mathcal{M} : E_\theta(x) = e\,\}\,,
  \label{eq:level-set}
\end{equation}
is a smooth embedded hypersurface of $\mathcal{M}$.
Geometrically, $\Sigma_e^{(\theta)}$ is the leaf of constant entanglement $e$, and $\nabla E_\theta(x)$ is everywhere normal to this leaf. In other words, if $v \in T_x \Sigma_e^{(\theta)}$ then:
\begin{equation}
  dE_\theta(x)[v] =
  g_{\text{FS}}(\nabla E_\theta(x),v)=0.
\end{equation}
In regions where all values of $E_\theta$ are regular, namely, $\nabla E_\theta(x)\neq 0$, the collection of level sets $\{\Sigma_e^{(\theta)}\}_e$ defines a foliation of $\mathcal{M}$ by constant-entanglement hypersurfaces. Then, the tangent space splits orthogonally as
\begin{equation}
  T_x\mathcal{M}
  \;=\;
  \mathrm{span}\{\nabla E_\theta(x)\}
  \;\oplus\;
  T_x\Sigma^{(\theta)}_{E_\theta(x)}\,,
  \label{eq:orth-split}
\end{equation}
where $\text{span}\{\cdot\}$ indicates the (one-dimensional) direction pointed out by the vector $\nabla E_{\theta}$ at the point $x\in\mathcal{M}$. Given any tangent vector $w\in T_x\mathcal{M}$, we can uniquely decompose it into a \textit{normal} (lying in $\mathrm{span}\{\nabla E_\theta(x)\}$) and a \textit{tangential} component (lying in $T_x\Sigma^{(\theta)}_{E_\theta(x)}$), namely:
\begin{equation}
  w = w^{\perp} + w^{\parallel},
\end{equation}
where
\begin{equation}
\begin{split}
      w^{\perp}&= \frac{g_{\text{FS}}(w,\nabla E_\theta(x))}
           {g_{\text{FS}}(\nabla E_\theta(x),\nabla E_\theta(x))}\,\nabla E_\theta(x),\\
  w^{\parallel} &= w - w^{\perp}\in T_x\Sigma^{(\theta)}_{E_\theta(x)}.
  \label{eq:proj-normal-tang}
\end{split}
\end{equation}
The normal component $w^{\perp}$ changes the value of $E_\theta$ to first order, while the tangential component $w^{\parallel}$ preserves it:
\begin{equation}
  dE_\theta(x)[w^{\parallel}] = 0,
  \qquad
  dE_\theta(x)[w^{\perp}]
  = g_{\text{FS}}(\nabla E_\theta(x),w^{\perp})\,.
\end{equation}
In particular, an infinitesimal displacement purely along $T_x\Sigma^{(\theta)}_{E_\theta(x)}$ corresponds to a change in the quantum state that leaves the entanglement $E_\theta$ unchanged, whereas motion with a component along $\nabla E_\theta$ (or $-\nabla E_\theta$) corresponds to an increase (or decrease) in entanglement to first order.

This orthogonal decomposition is the geometric backbone of what follows. 
Trajectories $\gamma(t)$ in projective Hilbert space are curves whose velocity $\dot{\gamma}(t)$ at each point can be resolved into a tangential part $\dot{\gamma}(t)^{\parallel}\in T_{\gamma(t)}\Sigma^{(\theta)}_{E_\theta(\gamma(t))}$ and a normal part $\dot{\gamma}(t)^{\perp}\propto \nabla E_\theta(\gamma(t))$.
In our coding scheme, we will interpret these two components as elementary \textit{tangential} and \textit{normal} moves with respect to the entanglement foliation.
In the companion geometric construction of entanglement entropy~\cite{DiCairanoGeoEntEntropy}, together with geometric tools built in Ref.~\cite{gori_configurational,di2021topology,di2022geometrictheory}, the same foliation supports a microcanonical-like measure based on the induced FS volume of the leaves and the density of states
\(
    \Omega_\theta(e)
\),
so that $E_\theta$ acquires an entropic meaning; however, this is left for a separate investigation. Here, we use this structure operationally: the leaves $\Sigma_e^{(\theta)}$ and the directions $\nabla E_\theta$ provide the geometric alphabet with which classical information is encoded in quantum trajectories.

\section{Geometric entanglement codes}

\subsection{Idea and basic structure}

We now formalize the intuitive idea.
A \emph{geometric entanglement code} consists, at the conceptual level, of:
\begin{itemize}
  \item a state manifold $(\mathcal{M},g_{\text{FS}})$;
  \item a family of entanglement functions $\{E_\theta\}_{\theta\in\Theta}$;
  \item a class of allowed unitary operations $\mathcal{U}\subset U(\mathcal{H})$;
  \item an encoding map that takes a classical message and a key $\theta$ and produces a sequence of unitary moves, hence a trajectory $\gamma$ in $\mathcal{M}$.
\end{itemize}

The informal rules are:
\begin{enumerate}
  \item The \emph{protocol} is public: everyone knows $(\mathcal{M},g_{\text{FS}})$, the family $\{E_\theta\}$, and the class of moves (unitaries) that may be used.
  \item The \emph{key} is the specific parameter $\theta\in\Theta$ that selects the actual entanglement functional and foliation relevant for encoding.
  \item The \emph{message} is encoded in the pattern with which a trajectory crosses the leaves of the foliation $\mathcal{F}_\theta = \{\Sigma_e^{(\theta)}\}$.
\end{enumerate}

In other words: the receiver who knows $\theta$ can interpret the evolution of entanglement along the trajectory and recover the message; an eavesdropper who does not know $\theta$ sees a sequence of states but cannot reconstruct the relevant ``vertical/tangential'' structure.

\subsection{Trajectories and elementary geometric moves}

Let $X_0\in\mathcal{M}$ be a fixed initial state (a ray in projective Hilbert space) where we use the notation $X_a:=|\psi_a\rangle$ for some label $a$. 
We consider trajectories generated by successive unitary operations,
\begin{equation}
  X_{k+1} = U_k \cdot X_k, 
  \qquad U_k\in\mathcal{U},\quad k=0,\dots,L-1,
\end{equation}
where $\mathcal{U}\subset U(\mathcal{H})$ is a chosen set of allowed unitaries.
The resulting discrete trajectory is the finite sequence
\begin{equation}
  \gamma = (X_0,X_1,\dots,X_L), \qquad X_k\in\mathcal{M}.
\end{equation}
One may think of $\gamma$ as a sampling of a piecewise-smooth curve $\gamma(t)$ at times $t_k$, with $X_k=\gamma(t_k)$; however, for our purposes, the discrete description is sufficient.

For a fixed entanglement functional $E_\theta$, the entanglement profile along the path is
\begin{equation}
  e_k^{(\theta)} := E_\theta(X_k), \qquad k=0,\dots,L.
\end{equation}
The elementary step $X_k\mapsto X_{k+1}$ is a small displacement on $(\mathcal{M},g_{\text{FS}})$.
Let $d_{\text{FS}}$ denote the FS distance, and define the FS step length
\begin{equation}
  \Delta s_k := d_{\text{FS}}(X_k,X_{k+1}),
\end{equation}
together with the entanglement increment
\begin{equation}
  \Delta e_k^{(\theta)} := e_{k+1}^{(\theta)} - e_k^{(\theta)}
  = E_\theta(X_{k+1}) - E_\theta(X_k).
\end{equation}
For sufficiently small steps, we may view $X_{k+1}$ as obtained from $X_k$ by an infinitesimal displacement $w_k\in T_{X_k}\mathcal{M}$, with
$\Delta s_k \approx \|w_k\|_{g_{\text{FS}}}$ and
\begin{equation}
  \Delta e_k^{(\theta)}
  \;\approx\;
  dE_\theta(X_k)[w_k]
  \;=\;
  g_{\text{FS}}\big(\nabla E_\theta(X_k), w_k\big),
  \label{eq:finite-diff-grad}
\end{equation}
where $\nabla E_\theta$ is the FS gradient introduced in Sec.~\ref{ssec:geometric-constr}.
Using the orthogonal splitting $T_{X_k}\mathcal{M} = \mathrm{span}\{\nabla E_\theta(X_k)\}\oplus T_{X_k}\Sigma_{E_\theta(x_k)}^{(\theta)}$, we can decompose
\begin{equation}
  w_k = w_k^{\perp} + w_k^{\parallel},
\end{equation}
with $w_k^{\perp}\propto \nabla E_\theta(X_k)$ and $w_k^{\parallel}\in T_{X_k}\Sigma_{E_\theta(X_k)}^{(\theta)}$.
Recall that, by construction, $w_k^{\parallel}$ is tangential to the constant-entanglement leaf and does not change $E_\theta$ to first order, while $w_k^{\perp}$ is normal to the leaf and is responsible for the variation of $E_\theta$.

In the discrete setting, we do not reconstruct $w_k$ explicitly but classify each step $X_k\to X_{k+1}$ \emph{operationally} from the observed pair $(\Delta s_k,\Delta e_k^{(\theta)})$.
Fix two small positive thresholds $\varepsilon_{\text{tan}}$ and $\varepsilon_{\text{norm}}$ with $\varepsilon_{\text{tan}} \ll \varepsilon_{\text{norm}}$.
We say that the move $X_k\to X_{k+1}$ is
\begin{itemize}
  \item approximately \emph{tangential} if
  \begin{equation}
    \frac{|\Delta e_k^{(\theta)}|}{\Delta s_k} < \varepsilon_{\text{tan}},
  \end{equation}
  i.e.\ the entanglement changes very little compared to the FS distance travelled, so the displacement is dominated by $w_k^{\parallel}$;
  \item approximately \emph{upward} if
  \begin{equation}
    \frac{\Delta e_k^{(\theta)}}{\Delta s_k} > \varepsilon_{\text{norm}},
  \end{equation}
  i.e.\ $E_\theta$ increases by an amount comparable to the step length, so the displacement has a significant component along $+\nabla E_\theta$;
  \item approximately \emph{downward} if
  \begin{equation}
    \frac{\Delta e_k^{(\theta)}}{\Delta s_k} < -\varepsilon_{\text{norm}},
  \end{equation}
  i.e.\ $E_\theta$ decreases and the displacement has a significant component along $-\nabla E_\theta$.
\end{itemize}
Steps that fall into the intermediate regime
$\varepsilon_{\text{tan}}\leq |\Delta e_k^{(\theta)}|/\Delta s_k \leq \varepsilon_{\text{norm}}$
can be treated according to any convenient convention (for instance rounded to the nearest of the three classes, or assigned to a separate ``mixed'' symbol if desired).

In this way, the foliation $\mathcal{F}_\theta$ and the FS geometry induce a simple \emph{alphabet of geometric moves},
\begin{equation}
  \mathcal{A}_{\text{geo}} = \{T,U,D\},
\end{equation}
where $T$ denotes a tangential step (approximately within a leaf $\Sigma^{(\theta)}_{E_\theta(X_k)}$), $U$ an upward step (increasing entanglement along $+\nabla E_\theta$), and $D$ a downward step (decreasing entanglement along $-\nabla E_\theta$).
A very simple coding strategy is then to discretize the evolution into $L$ steps and map each classical symbol (for instance a bit) to a choice of move in $\mathcal{A}_{\text{geo}}$ at each step, under constraints that keep the trajectory within a prescribed region of the manifold.
The resulting sequence of states $(X_0,\dots,X_L)$ is a path whose pattern of $T/U/D$ moves, evaluated with respect to the \emph{secret} entanglement functional $E_\theta$, carries the message.

\subsection{Public protocol vs.\ secret key}

From a cryptographic viewpoint, the crucial point is that the protocol itself must be considered public.
In our setting, this means:
\begin{itemize}
  \item the manifold $\mathcal{M}$ and its metric $g_{\text{FS}}$ are known;
  \item the functional form of the family $\{E_\theta\}$ is known, at least up to the parameters $\theta$;
  \item the set of allowed unitaries $\mathcal{U}$ and the mapping ``message $\to$ sequence of moves'' are known as algorithms.
\end{itemize}
The only thing that is not public is the actual value of $\theta$.
Equivalently, what remains secret is which foliation $\mathcal{F}_\theta$ is used to interpret the pattern of moves.

Eavesdroppers can, in principle, reconstruct the sequence of states $(x_k)$ if they can track the quantum evolution. However, without knowing $\theta$, they cannot determine which entanglement profile $e_k^{(\theta)}$ is relevant. Different choices of $\theta$ may lead to very different entanglement patterns for the same trajectory.

We stress that this is not yet a claim of security: to turn this into a serious cryptographic scheme, one should analyze how hard it is, given many trajectories, to infer $\theta$ or to recover the message without $\theta$.
Here, we do not attempt such an analysis which is left to future work. Instead, we build a minimal example that illustrates how the secret can be encoded as a hidden twist of the entanglement function.

\section{Toy example: two qubits with a twisted entanglement measure}

\subsection{Setup}

We now specialize to the simplest nontrivial case: two qubits.
The Hilbert space is
\begin{equation}
  \mathcal{H} = \mathbb{C}^2 \otimes \mathbb{C}^2, \qquad \mathcal{M} \simeq \mathbb{CP}^3\,,
\end{equation}
and we choose the computational basis $\{|00\rangle,|01\rangle,|10\rangle,|11\rangle\}$. The ``standard'' entanglement measure is the von Neumann entropy of the reduced state of subsystem $A$:
\begin{equation}
  E_{\text{std}}([\psi]) = -\mathrm{Tr}\Big(\rho_A \log \rho_A\Big), \quad \rho_A = \mathrm{Tr}_B|\psi\rangle\langle\psi|.
\end{equation}
For this toy model, we introduce a family of entanglement measures by composing $E_{\text{std}}$ with a local unitary on $A$:
\begin{equation}
  E_\theta([\psi]) := E_{\text{std}}\big([ W(\theta)\otimes\mathbb{I}|\psi\rangle ]\big).
\end{equation}
Here, $W(\theta)\in SU(2)$ is a single-qubit unitary that depends on a parameter $\theta$ (for instance, a rotation about some axis on the Bloch sphere).
Different choices of $\theta$ induce different foliations of $\mathcal{M}$ into constant-$E_\theta$ hypersurfaces.

The protocol is public in the sense that everyone knows that the entanglement measure is of this twisted form.
However, the actual value of $\theta$ used by the communicating parties is secret and serves as the key.

\subsection{A one-parameter family of states}

To construct a simple code, we restrict attention to a one-parameter family of states obtained by rotating the first qubit:
\begin{equation}
  |\psi_j\rangle = \big(R_y(j\delta)\otimes \mathbb{I}\big)|00\rangle, \qquad j=0,1,\dots,J,
\end{equation}
where $R_y(\alpha) = e^{-i\alpha\sigma_y/2}$ is a rotation about the $y$-axis on the Bloch sphere, and $\delta$ is a fixed angular step. Explicitly,
\begin{equation}
  |\psi_j\rangle = \cos\left(\frac{j\delta}{2}\right)|00\rangle
  + \sin\left(\frac{j\delta}{2}\right)|10\rangle.
\end{equation}
Note that all these states are product states with respect to the standard bipartition and the standard measure: $E_{\text{std}}([\psi_j]) = 0$ for all $j$.
From this point of view, they all lie on the same (trivial) entanglement leaf.

However, with respect to the twisted measure $E_\theta$, the situation is different:
\begin{equation}
  e_j^{(\theta)} := E_\theta([\psi_j]) 
  = E_{\text{std}}\big([ (W(\theta)R_y(j\delta)\otimes\mathbb{I})|00\rangle ]\big).
\end{equation}
For a fixed $\theta$, the function $j\mapsto e_j^{(\theta)}$ is in general nontrivial.
In particular, there will typically be ranges of $j$ where $e_j^{(\theta)}$ increases approximately monotonically, and ranges where it decreases.

The key point is:
\begin{itemize}
  \item The legitimate parties (Alice and Bob), who share the value of $\theta$, can compute or characterize the map $j\mapsto e_j^{(\theta)}$.
  \item An eavesdropper who does not know $\theta$ and only looks at $E_{\text{std}}$ sees $e_j^{\text{(std)}} = 0$ for all $j$, i.e., there is no entanglement structure at all.
\end{itemize}
From the eavesdropper's viewpoint, the trajectory remains on a single, featureless leaf.

\subsection{Elementary moves and encoding}

We define three elementary unitary transformations on $\mathcal{H}$:
\begin{align}
  U^+ &:= R_y(\delta)\otimes\mathbb{I}, \\
  U^- &:= R_y(-\delta)\otimes\mathbb{I}, \\
  T &:= \mathbb{I}\otimes\mathbb{I}.
\end{align}
Acting on the family $\{|\psi_j\rangle\}$, they implement the moves
\begin{align}
  U^+: &\quad |\psi_j\rangle \mapsto |\psi_{j+1}\rangle,\\
  U^-: &\quad |\psi_j\rangle \mapsto |\psi_{j-1}\rangle,\\
  T:   &\quad |\psi_j\rangle \mapsto |\psi_j\rangle.
\end{align}
We assume that the allowed indices $j$ are restricted to some finite interval where this simple picture holds.

Let the initial state be $|\psi_{j_0}\rangle$, with $j_0$ publicly known.
A classical bit string $b_1b_2\dots b_L$ can be encoded in a sequence of elementary moves as follows (one of many possibilities):
\begin{itemize}
  \item if $b_k = 1$, apply $U^+$ (``up'');
  \item if $b_k = 0$, apply $U^-$ (``down'').
\end{itemize}
After $L$ steps, the index is
\begin{equation}
  j_L = j_0 + \sum_{k=1}^{L} \sigma_k, 
  \qquad \sigma_k = \begin{cases}
    +1 & \text{if } b_k = 1,\\
    -1 & \text{if } b_k = 0.
  \end{cases}
\end{equation}
The corresponding trajectory in $\mathcal{M}$ is the sequence of rays $x_k = [\psi_{j_k}]$.

From the point of view of the twisted entanglement measure $E_\theta$, the path produces the profile
\begin{equation}
  e_k^{(\theta)} = E_\theta([\psi_{j_k}]), \quad k=0,\dots,L.
\end{equation}
By construction, each step $k\to k+1$ corresponds to an ``up'' move in the index $j$ if $b_k=1$, and to a ``down'' move if $b_k=0$.
Since Alice and Bob know $\theta$, they know the mapping $j\mapsto e_j^{(\theta)}$ and can reconstruct the bit string from the pattern of upward/downward steps in $E_\theta$.

\subsection{Decoding with and without the key}

Let us now distinguish the perspectives.

\paragraph*{Legitimate receiver.}
Bob knows $\theta$ and the initial index $j_0$.
He also has access, in principle, to the sequence of states or to the sequence of unitary transformations applied.
He can, therefore, reconstruct the indices $j_k$, compute the entanglement values $e_k^{(\theta)}$, and map each sign $j_{k+1}-j_k$ back to the bit $b_k$.
Equivalently, he can precompute and store the map $j\mapsto e_j^{(\theta)}$ and use it to interpret any trajectory built from the elementary moves.

\paragraph*{Eavesdropper.}
Eavesdroppers who do not know $\theta$ face a different situation. If they attempt to use the standard measure $E_{\text{std}}$, they find $E_{\text{std}}([\psi_{j_k}]) = 0$ for all $k$; i.e., there is no pattern at all.
If they know that the actual measure is of the twisted form $E_\theta$, they could, in principle, try to guess $\theta$ by tomography and parameter estimation.
However, without prior information on $\theta$, this becomes an instance of a ``hidden local unitary'' inference problem.

In this toy example, we do not claim that such an inference is hard. The point is simply that the key $\theta$ is encoded in the choice of foliation, and that the same physical trajectory may look completely structureless with respect to an inappropriate entanglement functional.

\subsection{A geometric BB84 toy protocol}

To illustrate how the present geometric framework can reproduce standard information--disturbance tradeoffs, it is useful to consider a minimal toy protocol in which our constructions reduce essentially to the well-known BB84 scheme~\cite{BennettBrassard84,BechmannPasquinucciPeres00,ShorPreskill00}.
For clarity, we work with a single qubit rather than a bipartite entangled system; in this case, the ``entanglement functional'' is replaced by a simple height function on the Bloch sphere. 
This toy model shows that the structure of \emph{secret foliations} and \emph{geometric moves} is flexible enough to encode the same kind of basis incompatibility that underlies BB84, even though a full security proof in higher-dimensional, genuinely entangled settings is left for future work.

\paragraph*{Geometry of the qubit as a projective manifold.}
For a qubit, the projective Hilbert space is
\(
\mathcal{M} \cong \mathbb{CP}^1 \cong \mathbb{S}^2
\),
and every pure state can be represented by a Bloch vector $\vec{r}=(x,y,z)$ with $|\vec{r}|=1$. 
The FS metric coincides (up to a constant factor) with the standard round metric on $\mathbb{S}^2$, and the FS distance $d_{\text{FS}}$ is proportional to the angle between Bloch vectors~\cite{BrodyHughston01,BengtssonZyczkowskiBook}.
Given any unit vector $\vec{n}\in \mathbb{S}^2$, we define the Pauli observable $\sigma_{\vec{n}} = \vec{n}\cdot\vec{\sigma}$ and the associated expectation value
\begin{equation}
  \langle \sigma_{\vec{n}} \rangle_\psi 
  := \langle \psi|\sigma_{\vec{n}}|\psi\rangle
  = \vec{n}\cdot\vec{r}.
\end{equation}
From this, we build a smooth function:
\begin{equation}
  E_{\vec{n}}(\psi) := \frac{1}{2}\big(1 + \langle \sigma_{\vec{n}} \rangle_\psi\big)
  = \frac{1}{2}(1+\vec{n}\cdot\vec{r}),
  \label{eq:En-qubit}
\end{equation}
which takes values in $[0,1]$ and can be viewed as a ``height function'' along the direction $\vec{n}$.
Its level sets $E_{\vec{n}}(\psi)=e$ are parallels on the Bloch sphere, and the FS gradient $\nabla E_{\vec{n}}$ points along $\vec{n}$: integral curves of $\nabla E_{\vec{n}}$ move ``upwards'' towards the pole defined by $\vec{n}$.

Within our notation, this is a special case of the general construction:
\begin{equation}
  E_\theta \;\equiv\; E_{\vec{n}_\theta}, \qquad
  \theta \mapsto \vec{n}_\theta\in S^2.
\end{equation}
The family $\{E_\theta\}$ induces a family of foliations of $\mathcal{M}$ by parallels with respect to different axes $\vec{n}_\theta$.
Even though $E_\theta$ is not an entanglement measure in this one-qubit example, its geometric role (height function, FS gradient, foliation by level sets) is identical to that of a genuine geometric entanglement functional in higher-dimensional settings~\cite{BengtssonZyczkowskiBook,BengtssonZyczkowskiIntro,HorodeckiRMP09,AnandanAharonov90,BraunsteinCaves94}.

\paragraph*{Two incompatible foliations: $\sigma_z$ and $\sigma_x$.}
We now select two particular directions on the Bloch sphere,
\begin{equation}
  \vec{n}_z = (0,0,1),
  \qquad
  \vec{n}_x = (1,0,0),
\end{equation}
and define
\begin{equation}
  E_z := E_{\vec{n}_z}, \qquad E_x := E_{\vec{n}_x}.
\end{equation}
The level sets of $E_z$ are the circles of constant $z$ (parallels with respect to the $z$-axis), while those of $E_x$ are circles of constant $x$ (parallels with respect to the $x$-axis).
The corresponding foliations $\mathcal{F}_{\sigma_z}$ and $\mathcal{F}_{\sigma_x}$ are maximally incompatible: their leaves intersect transversely, and their gradients point in orthogonal directions on the sphere (up to local rotations).
Formally, we may regard $\theta\in\{\sigma_z,\sigma_x\}$ as labeling two different choices of functional,
\(
    E_\theta \in \{E_z,E_x\}
\),
and therefore two different secret \textit{entanglement geometries}.

In this toy setting, the ``upward'' and ``downward'' geometric moves associated with $\theta=\sigma_z$ correspond to moving towards the north and south poles (eigenstates of $\sigma_z$), while those associated with $\theta=\sigma_x$ move towards the eigenstates of $\sigma_x$.
Thus, the two foliations $\mathcal{F}_{\sigma_z}$ and $\mathcal{F}_{\sigma_x}$ geometrically encode the two mutually unbiased bases used in the BB84 protocol~\cite{BennettBrassard84,BechmannPasquinucciPeres00}.

\paragraph*{Encoding bits as geometric moves.}
We now specify a one-step protocol that can be iterated over multiple rounds.
Fix a reference state $X_0\in\mathcal{M}$, for instance, the north pole corresponding to $|0\rangle$.
Each use of the channel proceeds as follows:
\begin{enumerate}
  \item Alice chooses a basis label $\theta\in\{Z,X\}$ uniformly at random. This is the ``geometric key'' that selects which foliation $E_\theta$ is relevant in that round.
  \item Alice chooses a bit $b\in\{0,1\}$ to encode.
  \item Conditioned on $(\theta,b)$, Alice applies a unitary transformation $U_{\theta,b}$ that moves the state $x_0$ either ``upwards'' or ``downwards'' with respect to the foliation defined by $E_\theta$:
  \begin{equation}
    X_1 = U_{\theta,b}\cdot X_0,
    \qquad
    x_1\in\mathcal{M}.
  \end{equation}
  For concreteness, one may take $\theta = \sigma_z$:
  \begin{align}
      b=0 \Rightarrow X_1 = |0\rangle,\quad
      b=1 \Rightarrow X_1 = |1\rangle; 
    \end{align}
     or $\theta = \sigma_x$, then:
    \begin{align}
      b=0 \Rightarrow X_1 = |+\rangle,\quad
      b=1 \Rightarrow X_1 = |-\rangle,
  \end{align}
  where $|0\rangle,|1\rangle$ are the eigenstates of $\sigma_z$ and $|\pm\rangle$ the eigenstates of $\sigma_x$.
  In terms of~\eqref{eq:En-qubit}, these choices correspond to
  \begin{align}
    E_Z(|0\rangle)&=1,\; E_Z(|1\rangle)=0,
   \\
    E_X(|+\rangle)&=1,\; E_X(|-\rangle)=0,
  \end{align}
  so $b$ is precisely the ``up'' (high $E_\theta$) or ``down'' (low $E_\theta$) label with respect to the chosen foliation.
\end{enumerate}
In this language, each round is a trajectory $\gamma=(X_0,X_1)$ of length $L=1$, and the message is encoded in a single geometric move $U$ or $D$ relative to the foliation $\mathcal{F}_\theta$.
The set of possible states $\{X_1\}$ is exactly the usual BB84 signal set $\{|0\rangle,|1\rangle,|+\rangle,|-\rangle\}$.

\paragraph*{Decoding with and without the key.}
Bob receives the state $x_1$ and must infer $b$.
If he shares the key $\theta$ with Alice, he can measure in the corresponding eigenbasis:
\begin{itemize}
  \item if $\theta=\sigma_z$, Bob measures $\sigma_z$ and decodes
    $b=0$ for outcome $|0\rangle$ (upwards with respect to $E_Z$),
    $b=1$ for outcome $|1\rangle$ (downwards);
  \item if $\theta=\sigma_x$, Bob measures $\sigma_x$ and decodes
    $b=0$ for outcome $|+\rangle$ (upwards with respect to $E_X$),
    $b=1$ for outcome $|-\rangle$ (downwards).
\end{itemize}
From the geometric viewpoint, Bob is estimating whether the single step $\gamma=(X_0,X_1)$ has moved along $+\nabla E_\theta$ or $-\nabla E_\theta$, i.e., whether the entanglement functional $E_\theta$ (here simply an expectation value) has increased or decreased.

An eavesdropper, Eve, who does not know $\theta$, faces the usual BB84 dilemma: the four possible states lie on two incompatible foliations, and there is no single POVM that simultaneously distinguishes them all without error and without disturbing them~\cite{ScaraniRMP09,HorodeckiRMP09,ShorPreskill00}.
If Eve measures in the $Z$-basis when $\theta=\sigma_x$, or vice versa, she induces random outcomes from her perspective and, more importantly, disturbs the state so that Bob's later measurement statistics exhibit an increased error rate.
This is precisely the information--disturbance mechanism that underlies BB84 security proofs~\cite{ScaraniRMP09,ShorPreskill00}.

\paragraph*{Interpretation within the general framework.}
This toy protocol shows that our geometric language can reproduce textbook QKD already in the simplest case:
\begin{itemize}
  \item the family of functionals $\{E_\theta\}$ plays the role of \emph{basis choice};
  \item each $E_\theta$ defines a foliation $\mathcal{F}_\theta$ and a gradient flow $\pm\nabla E_\theta$;
  \item the classical bit $b$ is encoded as an ``up'' or ``down'' geometric move along $\pm\nabla E_\theta$;
  \item the key $\theta$ tells Bob which foliation to use when interpreting the move.
\end{itemize}
In higher-dimensional and genuinely entangled settings, $E_\theta$ will be a geometric entanglement functional on projective Hilbert space, and the foliations will be nontrivial hypersurfaces of constant entanglement.
The qubit example above suggests that by choosing a family $\{E_\theta\}$ associated with incompatible observables, and by encoding messages as patterns of normal and tangential moves relative to these foliations, one can engineer protocols where ignorance of $\theta$ forces any eavesdropper to perform measurements that disturb the trajectory in a detectable way.
A quantitative security analysis along the lines of Shor-Preskill~\cite{ShorPreskill00} and related work is beyond the scope of this first exploration, but the toy model already clarifies how ``secret entanglement geometry'' can act as a cryptographic resource rather than a purely descriptive tool.

\section{Discussion and outlook}

In summary, we have proposed treating \emph{entanglement geometry} as a cryptographic resource. On top of a fully public kinematic structure—the projective Hilbert space $(\mathcal{M},g_{\text{FS}})$ and its Fubini-Study geometry—we introduce a family of entanglement functionals $\{E_\theta\}_{\theta\in\Theta}$ and use the \emph{choice of foliation} $\mathcal{F}_\theta$ as (part of) the secret key. In this setting, classical information is not only encoded in which quantum states are prepared but also in how trajectories move across constant-entanglement leaves, as seen through the “correct’’ height function $E_\theta$. The same path in state space can therefore carry a readable message for a party who knows $\theta$ and appear almost featureless to an observer who does not.

We formalized this idea in terms of \emph{geometric entanglement codes} and illustrated it with two minimal constructions: a two-qubit toy model in which the key is a local unitary that twists the entanglement measure, and a geometric reformulation of BB84 where incompatible foliations of projective space play the role of mutually unbiased bases. These examples show that \textit{secret entanglement geometry} can reproduce standard information-disturbance tradeoffs and offer a new language in which to organize them. 

The natural next steps are to move beyond toy models: to analyze the learnability of $\theta$ from data, to incorporate noise and mixed states, and to design families of functionals $E_\theta$ whose foliations are intrinsically incompatible in higher-dimensional and multipartite settings. More broadly, the framework suggests that geometric tools originally developed to \emph{describe} quantum states and entanglement can be repurposed to \emph{engineer} protocols where what is secret is not only a basis or a key string but also a method for reading the landscape of entanglement.

\end{document}